\documentclass[english]{article}
\usepackage[T1]{fontenc}
\usepackage[utf8]{inputenc}
\usepackage{float}
\usepackage{amsmath}
\usepackage{graphicx}
\usepackage{esint}

\makeatletter
\newcommand{\lyxaddress}[1]{
	\par {\raggedright #1
	\vspace{1.4em}
	\noindent\par}
}

\usepackage{babel}
\usepackage[superscript,biblabel]{cite}

\makeatother

\usepackage{babel}
\begin{document}
\title{Mesoscopic Correlations in Aqueous Alkylamine Mixtures Between Molecular
and Micro Emulsions}
\author{Aurélien Perera \thanks{aup@lptmc.jussieu.fr}}
\maketitle

\lyxaddress{Laboratoire de Physique Théorique de la Matière Condensée (UMR CNRS
7600), Sorbonne Université, 4 Place Jussieu, F75252, Paris cedex 05,
France.}
\begin{abstract}
Understanding how molecular correlations give rise to mesoscale organization
is central to the physics of complex fluids such as hydrogen-bonded
mixtures. In this work, we develop a mesoscale bridge formalism that
connects the site--site Ornstein--Zernike (SSOZ) framework to the
field theoretical Teubner-Strey (TS) approach. This bridge highlights
how local orientational correlations, typically lost in the SSOZ closure,
reemerge as effective long-range components at the mesoscale. The
resulting theory provides a unified description of density fluctuations
spanning molecular to mesoscopic length scales. The approach is illustrated
using X-ray scattering spectra from simulated and experimental hydrogen-bonded
fluids, showing that the TS representation captures the essential
features of the mesoscale structure. Beyond this specific application,
the proposed formalism offers a general route to interpret the structural
crossover between microscopic interactions and collective mesoscale
organization in complex fluids, including aqueous and amphiphilic
systems. 
\end{abstract}

\section{Introduction\protect}\label{sec:Introduction}

Liquids can be broadly classified into two main categories, separated
by a wide conceptual ``interface'': simple liquids\cite{Textbook_Hansen_McDonald}
and organized liquids\cite{deGennes-softmatter}. Typical examples
of the first class include for instance liquid nitrogen, benzene,
or weakly polar cesium sulfide, where interactions remain essentially
isotropic and short-ranged\cite{simpleLiquidJeppe}. The second class
encompasses soft-matter and biological liquids, characterized by complex
molecular architectures maintained by hard covalent bonds and soft
bonds such as hydrogen bonds. 

In between these two classes, \textit{molecular emulsions} \cite{AupPAC,AupDomainOrdering}
are characterized by partial microscopic segregation, leading to transient,
nanoscale domains stabilized by molecular association through hydrogen
bonding and/or hydrophobic effects. Because these systems combine
structural disorder with local organization, one may introduce it
as a third broad category of liquids, represented by water and aqueous
mixtures. Despite the extensive literature devoted to each of these
three families, there has been little theoretical effort to unify
them under a single conceptual framework. Herein, this unification
is attempted, guided by fundamental ideas from statistical physics.
This approach is illustrated using aqueous alkylamine mixtures as
model systems, since they naturally encompass all three categories
within a single family.

This ambition echoes two historical viewpoints. The physicist Lev
Landau, as cited by de Gennes,\cite{deGennesDianoux} regarded the
theories of liquids as “neither convincing nor useful” in contrast
with that of gases and solids---a challenge that de Gennes later
addressed in his pioneering work on soft matter.\cite{deGennes-softmatter}.
On the other hand, Rowlinson and Swinton\cite{rowlinson-swinton}
famously remarked that aqueous mixtures of non-electrolytes would
probably be the last liquids to be fully understood. These complementary
perspectives---from a theoretical physicist, a soft-matter pioneer,
and two chemists---motivate the search for a unifying view of the
liquid state, encompassing simple disorder with organized disorder.
They are also a good starting point to understand why microscopic
liquid state theories are unable to reach beyond simple liquids, while
field theories are not meant to the study of small molecules in solution.
Molecular emulsions, such as aqueous amine mixtures are the strategic
meeting place where both approches fail. Our hypothesis is that such
systems encompass both large scale fluctuations and organized structures,
while still remaining on the side of molecular liquids. In that, they
require consistent theories of fluctuations, while staying close to
the molecular level.

The key physical concept borrowed from fundamental theory is the competition
between fluctuations and stability, which underlies the physics of
critical phenomena. In such theories, fluctuations---particularly
in density or concentration---are associated with $k=0$ modes, and
the divergence of the corresponding response function signals proximity
to a critical point. Some liquids, such as unoriented nematics, cannot
be reduced to a single state point but rather form a continuum of
critical states exhibiting permanent critical opalescence\cite{deGennesLiqCryst},
but which are outside the scope of our present concern. Below this
context is examined for aqueous alkylamines as opposed to aqueous
alcohols.

In binary mixtures, phase separation typically follows a binodal with
an upper critical solution temperature (UCST). At high temperature,
miscibility is ensured by thermal disorder, while at lower temperatures
the antagonistic interactions between unlike species dominate, leading
to demixing. This is the case for aqueous alcohols, which demix beyond
1-propanol. The antagonism arises between charged polar groups (water
and hydroxyl) and neutral alkyl groups, whose influence grows with
increasing chain length and eventually overcomes the Coulomb affinity
between the polar sites. Conversely, systems exhibiting a lower critical
solution temperature (LCST) display the opposite behavior: they demix
at high temperature but remix upon cooling. Classical examples include
aqueous mixtures of poly(N-isopropylacrylamide) (PNIPAM), poly(ethylene
oxide) (PEO), and tertiary butanol, among others. Aqueous alkylamines
belong to this latter category, showing a clear LCST beyond hexylamine---that
is, for sufficiently long alkyl chains. One may then ask what microscopic
mechanism allows the Coulomb association between water and the amine
headgroup to promote remixing at lower temperature, despite the reduced
thermal agitation. Walker and Vause\cite{WalkerVause} proposed such
a scenario within a renormalization group theory (RGT) framework,
where tight solute--solvent dimer formation drives the LCST. In their
picture, the homogeneous liquid corresponds to a dense fluid of such
dimers.

Our results \cite{our_expt_amin,our_simu_amin} confirm this scenario
only partially. The formation of water--amine dimers through N--H···O
hydrogen bonds is indeed observed, but the resulting liquid is not
homogeneous. Instead, it is micro-heterogeneous, consisting of water
nano-domains whose surfaces are saturated by amine nitrogen atoms.
In other words, aqueous amines do not form a uniform liquid of dimers,
but a disordered network of interfacial domains---an aspect entirely
missed by the WV approach. This shortcoming of a very successful and
leading fundamental theory can be interpreted as a need of a more
microscopic approach which accounts for the observed features.

The alternative paradigm proposed herein is the formation of nano-domains,
conceptually similar to the structure of micro-emulsions. The question
examined here is if this picture is compatible with the Teubner--Strey
(TS) model of modulated phases\cite{EXP_Teubner_Strey}, which describes
the shift of the instability from $k=0$ to a finite $k\neq0$. Such
a transfer is typically associated with mesocopic/macroscopic ordering,
as observed for micelle formation, lamellar or other layered phases.\cite{SafranBook}
In the case of aqueous amines, the observed structure corresponds
instead to a disordered melt of broken layers---reorganized water
and alkyl nano-domains forming what we have previously call a \textit{domain-ordered}
liquid\cite{AupDomainOrdering}. At low amine concentrations, when
there are insufficient amine--water dimers to stabilize extended
water regions, macroscopic phase separation occurs, consistent with
the LCST behavior. As the amine concentration increases, the water
domains shrink and their surfaces become saturated with amine headgroups.
The domains then decrease in size until, at high amine content, the
system becomes a continuous amine phase embedding small water clusters. 

The next section examines how the Teubner-Strey theory emerges from
a field theorectic approach and underline the steps which are recovered
in the third section where a more microscopic approach is developed,
specifically introduce the conjecture on the meso-ranged bridge in
order to reconcile both approaches. In the fourth section illustrates
this progression using aqueous hexylamine and octylamine mixtures,
highlighting the correspondence with the Teubner--Strey description
inferred from molecular dynamics simulations. The last two sections
gather remarks about the developments presented herein, and the final
conclusion.

\section{The Teubner--Strey behaviour of aqueous octylamine mixtures}

In their seminal 1987 paper, Teubner and Strey derived a compact expression
for the small- angle scattering intensity starting from a classical
Landau-type free energy with a local order parameter $\psi$: 
\begin{equation}
\beta F_{\mathrm{FT}}[\psi]=\int_{\Lambda}f(\psi,\nabla\psi,\Delta\psi)\,d\mathbf{r},\label{FLandau}
\end{equation}
where $\beta=1/k_{B}T$ and the subscript FT stands for ``field theory.''
The free-energy density $f$ is generally written in the Landau--de
Gennes form, including powers of $\psi$ and its spatial derivatives:
\begin{equation}
f=\sum_{n=0}a_{n}\psi^{n}+\sum_{n=1}c_{n}(\nabla^{2n}\psi).\label{fLdG}
\end{equation}
An important aspect of this formulation is the short-distance cutoff
parameter $\Lambda$, which indicates that both the integrand and
the resulting free energy are valid only for distances larger than
$\Lambda$ (in lattice theories, $\Lambda$ would correspond to the
lattice spacing). This cutoff acknowledges that the microscopic physics
below $\Lambda$ is not described by the model. In quantum field theory,
for instance, this missing physics concerns the structure of the vacuum,
for which no explicit theory exists. Approaching the $\Lambda\to0$
limit leads to the well-known ultraviolet divergences, usually handled
through renormalization procedures---an indirect way to bypass the
unknown short-range physics. In classical, off-lattice field theories,
$\Lambda$ is often formally set to zero, and the analysis focuses
instead on the large-$r$ (infrared) or small-$k$ limit, as in the
renormalization group treatment of critical points\cite{MaBook}.
Depending on the values and signs of the coefficients $a_{n}$ and
$c_{n}$, several scenarios arise for classical criticality, such
as the Ornstein--Zernike (OZ) form\cite{BernePecora}, Cahn--Hilliard
spinodal decomposition\cite{CahnHilliard}, Lifshitz points, and other
modulated phases\cite{Lifshitz}. A particularly relevant case is
defined by $a_{n}=0$ except for $a_{2}>0$, $c_{1}<0$, and $c_{2}>0$,
which describes the characteristic features of micro-emulsions, i.e.,
systems with a negative “microscopic” surface tension. Under these
conditions, the free energy reduces to: 
\begin{equation}
\beta F_{\mathrm{FT}}=\int\left[a_{2}\psi^{2}+c_{1}(\nabla\psi)^{2}+c_{2}(\Delta\psi)^{2}\right]d\mathbf{r}.\label{FTS}
\end{equation}
The usual approach consists in expanding $F_{\mathrm{FT}}$ around
the equilibrium mean value $\bar{\psi}$ by setting $\psi=\bar{\psi}+\delta\psi$
and truncating the expansion at quadratic order: 
\begin{equation}
\beta F_{\mathrm{FT}}[\psi]=\beta F_{\mathrm{FT}}[\bar{\psi}]+\frac{1}{2}\int d\mathbf{r}\int d\mathbf{r}'\,\delta\psi(\mathbf{r})\Gamma(|\mathbf{r}-\mathbf{r}'|)\delta\psi(\mathbf{r}'),\label{FT-expan2}
\end{equation}
where the linear term vanishes by virtue of the equilibrium condition
\[
\frac{\delta\beta F_{\mathrm{FT}}[\psi]}{\delta\psi(1)}\Big|_{eq}=0,
\]
and the kernel $\Gamma(|\mathbf{r}-\mathbf{r}'|)$ is the second functional
derivative of the free energy, 
\begin{equation}
\Gamma(|\mathbf{r}-\mathbf{r}'|)=\frac{\delta^{2}\beta F_{\mathrm{FT}}[\psi]}{\delta\psi(\mathbf{r})\,\delta\psi(\mathbf{r}')}\Big|_{eq}.\label{Gamma}
\end{equation}
The generating functional for zero external field is the partition
function, 
\begin{equation}
Z=\int D\psi\,\exp[-\beta F_{\mathrm{FT}}[\psi]]\sim\int D\psi\,\exp\left[-\frac{1}{2}\int d\mathbf{r}\int d\mathbf{r}'\,\delta\psi(\mathbf{r})\Gamma(|\mathbf{r}-\mathbf{r}'|)\delta\psi(\mathbf{r}')\right],\label{Z}
\end{equation}
which is a continuous Gaussian functional integral. In Fourier space
this becomes: 
\begin{equation}
Z\sim\int d\psi\,\exp\left[-\frac{1}{2}\int d\mathbf{k}\int d\mathbf{k}'\,\delta\psi(\mathbf{k})\Gamma(|\mathbf{k}+\mathbf{k}'|)\delta\psi(-\mathbf{k}')\right],\label{Z-FT}
\end{equation}
where the kernel is readily obtained from Eq.~(\ref{FTS}) as 
\begin{equation}
\Gamma(k)=a_{2}+c_{1}k^{2}+c_{2}k^{4}.\label{Gamma(k)}
\end{equation}
The scattering intensity $I(k)$ is proportional to the structure
factor $S(k)=\langle\psi(\mathbf{k})\psi(-\mathbf{k})\rangle$, which
in the Gaussian approximation is simply the variance of the fluctuations:
\[
\langle\psi(\mathbf{k})\psi(-\mathbf{k})\rangle\propto\Gamma^{-1}(k).
\]
This definition leads directly to the Teubner--Strey (TS) expression:
\begin{equation}
I(k)\propto\frac{A}{a_{2}+c_{1}k^{2}+c_{2}k^{4}},\label{I-TS}
\end{equation}
where $A$ is a normalization constant. The TS form contrasts with
the Ornstein--Zernike limit, which corresponds to $c_{1}>0$ and
$c_{2}=0$. In the following sections, this formalism will be rederived
from a more microscopic approach. Teubner and Strey demonstrated that
Eq.~(\ref{I-TS}) successfully reproduces small-angle neutron scattering
(SANS) data for a variety of micro-emulsions, such as D$_{2}$O--decane--
36/58 AOT droplet systems\cite{EXP_Teubner_Strey}. Equation~(\ref{I-TS})
also implies a corresponding real-space correlation function of the
form: 
\begin{equation}
I(r)=\bar{d}\,\frac{\exp(-r/\xi)}{r}\,\sin\!\left(\frac{r}{\bar{d}}\right),\label{I(r)}
\end{equation}
where the oscillation period is governed by the characteristic domain
size $d=2\pi\bar{d}$, and the correlation length $\xi$ and domain
spacing $d$ are related to the parameters in Eq.~(\ref{I-TS}) by\cite{EXP_Teubner_Strey}:
\begin{align}
d & =2\pi\left[\frac{1}{2}\sqrt{\frac{a_{2}}{c_{2}}}-\frac{1}{4}\frac{c_{1}}{c_{2}}\right],\label{d-TS}\\
\xi & =\left[\frac{1}{2}\sqrt{\frac{a_{2}}{c_{2}}}+\frac{1}{4}\frac{c_{1}}{c_{2}}\right].\label{xi-TS}
\end{align}
The correlations represented by $I(r)$ therefore decay exponentially
with correlation length $\xi$, as in the Ornstein--Zernike case,\cite{Textbook_Hansen_McDonald}
but include an additional oscillatory term whose period is determined
by the domain size $d$. 

To connect this mesoscopic description with a microscopic one, one
can relate the real-space correlations to the atom--atom pair correlation
functions $g_{ij}(r)$ through the total atom-- atom structure factors
$S_{ij}(k)$ using the Debye relation:\cite{Debye1,Debye2} 
\begin{equation}
I(k)=r_{0}^{2}\rho\sum_{ij}f_{i}(k)f_{j}(k)S_{ij}(k),\label{Ik-1}
\end{equation}
where $\rho=N/V$ is the molecular number density, $f_{i}(k)$ are
the atomic form factors, and $r_{0}=2.8179\times10^{-13}$ cm is the
classical electron radius. The total structure factor $S_{ij}(k)$
is defined as:\cite{Pozar2020} 
\begin{equation}
S_{ij}(k)=w_{ij}(k)+\rho H_{ij}(k),\label{ST(k)-1}
\end{equation}
where $w_{ij}(k)$ is the intra-molecular structure factor and $H_{ij}(k)$
is related to the Fourier transform of the intermolecular pair correlation
function: 
\begin{equation}
H_{ij}(k)=\int d\mathbf{r}\,\left[g_{ij}(r)-1\right]e^{i\mathbf{k}\cdot\mathbf{r}}.\label{Sk-1}
\end{equation}
The purpose of the section is to reconcile Eqs.~(\ref{I-TS}) and
(\ref{Ik-1}) by showing that the former can be derived from the latter
in the small-$k$ limit, and to illustrate the physical relevance
of this correspondence for aqueous alkylamine mixtures.

\section{Liquid-state statistical theory basis for TS behaviour\protect}\label{sec:Liquid-state-statistical}

Soft-matter systems are often described by field-theoretic, mesoscopic
approaches in which some atomic details (e.g., explicit solvent degrees
of freedom) are coarse-grained\cite{CoarseGrainSoftMatter,LevittNobelcoarseGrain}.
At the opposite end, molecular liquids admit a fully microscopic statistical
description\cite{GrayGubbins}. A key ingredient of the field-theoretic
route is a coarse-grained order parameter (a random field) from which
mesoscopic correlation functions and structure factors---such as
Eq.~(\ref{I-TS})---are built. Liquid-state theory also relies on
random variables, but at the microscopic level: the site (atom) density
\begin{equation}
\rho_{i_{a}}(\mathbf{r})\equiv\sum_{n\in i_{a}}\delta\!\left(\mathbf{r}-\mathbf{r}_{n}\right),\label{rho_ia}
\end{equation}
where $i_{a}$ labels site $i$ on species $a$ (e.g.\ $i=\mathrm{O_{W}}$
and $a=\mathrm{H_{2}O}$, or $i=\mathrm{N}$ for $a=\mathrm{amine}$),
and the sum runs over all atoms of type $i_{a}$. From these microscopic
variables, one constructs various correlation functions, and in particular
the microscopic pair correlation functions $g_{i_{a}j_{b}}(r)$. Our
aim is to relate the mesoscopic and microscopic descriptions on a
single, controlled footing. Aqueous alkylamines provide an ideal testbed,
because the relevant length scales (molecular vs.\ domain) are both
experimentally accessible and well-separated but coupled.

Theoretical developments below show how the two routes connect. The
approach departs from the standard literature by identifying the \emph{bridge
function} as the natural carrier of mesoscopic (domain-scale) oscillations.

\subsection{Statistical theory of molecular emulsions}

Consider molecular liquids as atomistic fluids in which molecular
connectivity acts as a constraint. Each site is labeled by the composite
index $i_{a}$, with $a$ the species index (water, amine, …) and
$i$ the site identity within that species (O, H, N, C$_{1}$, …).
The site--site pair distribution satisfies the exact Morita--Hiroike
relation 
\begin{equation}
g_{i_{a}j_{b}}(r)\;=\;\exp\!\Big[-\beta v_{i_{a}j_{b}}(r)\;+\;h_{i_{a}j_{b}}(r)\;-\;c_{i_{a}j_{b}}(r)\;+\;b_{i_{a}j_{b}}(r)\Big],\label{g12}
\end{equation}
where $v_{i_{a}j_{b}}(r)$ is the pair potential between sites $i_{a}$
and $j_{b}$ at distance $r$, $h_{i_{a}j_{b}}(r)=g_{i_{a}j_{b}}(r)-1$,
$c_{i_{a}j_{b}}(r)$ is the direct correlation function, and $b_{i_{a}j_{b}}(r)$
is the \emph{bridge function}. The latter is an infinite series over
higher-order direct correlation functions, 
\begin{equation}
b_{i_{a}j_{b}}(r)\;=\;\sum_{n\ge2}\frac{1}{n!}\sum_{\{\ell_{c_{k}}^{(k)}\}_{k=3}^{n+1}}\;\prod_{k=3}^{n+1}\rho_{\ell_{c_{k}}^{(k)}}\;\int\!\prod_{k=3}^{n+1}d\mathbf{r}_{k}\;\Bigg[\prod_{k=3}^{n+1}h_{i_{a}\,\ell_{c_{k}}^{(k)}}(r_{1k})\Bigg]\,c_{\{\ell_{c_{k}}^{(k)}\}\,j_{b}}^{(n+1)}(\mathbf{r}_{13},\ldots,\mathbf{r}_{1\,n+1}),\label{b12}
\end{equation}
where $c^{(n+1)}$ denotes the $(n\!+\!1)$-body direct correlation
function and the set $\{\ell_{c_{k}}^{(k)}\}$ runs over $n-1$ intermediate
sites. To illustrate this expression with rather heavy indexing conventions,
the explicit expression for the first term is given:
\[
b_{i_{a}j_{b}}(r)=\sum_{l_{c},m_{d}}\frac{\rho_{l_{c}}\rho_{m_{d}}}{2!}\int d\mathbf{r}_{3}\int d\mathbf{r}_{4}h_{i_{a}l_{c}}(r_{13})h_{i_{a}m_{d}}(r_{13})c_{l_{c}m_{d}j_{b}}^{(3)}(r_{13},r_{23})
\]

\noindent From the presence of all order direct correlations, it is
evident that solving Eq.~(\ref{b12}) exactly is intractable, and
even the convergence of the series is subtle (lack of convergence
is known in 1D \cite{AUP_BFMT}; indications exist of delicate behavior
in 3D \cite{BarratHansenPastore}. This motivates an alternative route
guided by asymptotics and physical separation of length scales.

In the large-$r$ limit, expanding the exponential in Eq.~(\ref{g12})
to first order yields an exact asymptotic relation for the direct
correlation function: 
\begin{equation}
c_{i_{a}j_{b}}(r\to\infty)\;=\;-\beta\,v_{i_{a}j_{b}}(r)\;+\;b_{i_{a}j_{b}}(r).\label{c-asym}
\end{equation}
This is a key identity, which emphasizes a point often overlooked:
while $b_{i_{a}j_{b}}(r)$ is usually \emph{assumed} shorter-ranged
than dispersion forces (and therefore neglected in HNC-like closures),
complex liquids with self-assembly at multiple scales (soft and biological
matter) generally require a \emph{mesoscopic}, long-distance contribution
in $b_{i_{a}j_{b}}(r)$ to encode domain order.

The direct correlation is therefore decomposed as 
\begin{equation}
c_{i_{a}j_{b}}(r)\;=\;c_{i_{a}j_{b}}^{(R)}(r)\;+\;b_{i_{a}j_{b}}^{(SR)}(r)\;+\;b_{i_{a}j_{b}}^{(MR)}(r),\label{cBMR}
\end{equation}
where $c^{(R)}$ is a reference (short-ranged) contribution (e.g.\ from
an integral-equation closure), $b^{(SR)}$ is a short-ranged bridge
correction (often tuned to match thermodynamic targets such as internal
energy or compressibility), and $b^{(MR)}$ captures the essential
mesoscopic contribution responsible for domain-scale correlations:
\begin{equation}
b_{i_{a}j_{b}}^{(MR)}(r\to\infty)\;=\;b_{i_{a}j_{b}}(r).\label{BMR}
\end{equation}

Next, weak microscopic inhomogeneities around a uniform state are
considered, with local densities given by Eq.~(\ref{rho_ia}) and
\begin{equation}
\delta\rho_{i_{a}}(\mathbf{r})\;\equiv\;\rho_{i_{a}}(\mathbf{r})-\rho_{i_{a}}.\label{delta-rho}
\end{equation}
The \emph{total} direct correlation functions are generated by functional
derivatives of the Helmholtz free-energy with respect to the site
densities: 
\begin{equation}
\mathcal{C}_{\{j_{b}:n\}}^{(n)}(\mathbf{r}_{1},\ldots,\mathbf{r}_{n})\;=\;\frac{\delta^{(n)}\beta F[\{\rho_{i_{a}}\}]}{\prod_{\{j_{b}:n\}}\prod_{m=1}^{n}\delta\rho_{j_{b}}(\mathbf{r}_{m})}.\label{Tdcfn}
\end{equation}
Expanding $F[\{\rho_{i_{a}}\}]$ around the homogeneous equilibrium
($\beta F_{0}$) yields 
\begin{align}
\beta F[\{\rho_{i_{a}}\}] & =\beta F_{0}+\int d1\sum_{i_{a}}\left.\frac{\delta\beta F[\{\rho_{\alpha}\}]}{\delta\rho_{i_{a}}(1)}\right|_{0}\delta\rho_{i_{a}}(1)\nonumber \\
 & \quad+\frac{1}{2}\int d1\int d2\sum_{i_{a}j_{b}}\left.\frac{\delta^{2}\beta F[\{\rho_{i_{a}}\}]}{\delta\rho_{i_{a}}(1)\,\delta\rho_{j_{b}}(2)}\right|_{0}\delta\rho_{i_{a}}(1)\,\delta\rho_{j_{b}}(2)+\ldots,\label{F-series}
\end{align}
where derivatives are evaluated at uniform densities (subscript $0$).
The linear term vanishes by equilibrium: 
\begin{equation}
\left.\frac{\delta\beta F[\{\rho_{\alpha}\}]}{\delta\rho_{i_{a}}(1)}\right|_{0}=0.\label{dF/drho=00003D0}
\end{equation}
Recognizing the second functional derivative as the total direct correlation
function $\mathcal{C}_{i_{a}j_{b}}(\mathbf{r}_{1},\mathbf{r}_{2})$
and truncating at quadratic order leads to 
\begin{equation}
\beta F[\{\rho_{i_{a}}\}]\;=\;\beta F_{0}\;-\;\frac{1}{2}\int d\mathbf{r}\int d\mathbf{r}'\;\boldsymbol{\delta\rho}(\mathbf{r})\,\mathbf{C}(|\mathbf{r}-\mathbf{r}'|)\,\boldsymbol{\delta\rho}(\mathbf{r}'),\label{F-IET}
\end{equation}
where $\boldsymbol{\delta\rho}(\mathbf{r})$ collects all site-density
deviations and $\mathbf{C}$ is the matrix of total site--site direct
correlation functions. This expression mirrors the field-theory quadratic
form in Eq.~(\ref{FT-expan2}), but here it is derived \emph{microscopically}.

The matrix elements can be written as 
\begin{equation}
\mathcal{C}_{i_{a}j_{b}}(|\mathbf{r}-\mathbf{r}'|)=\sqrt{\rho_{i_{a}}\rho_{j_{b}}}\Big[\,V_{i_{a}j_{b}}(|\mathbf{r}-\mathbf{r}'|)-\sqrt{\rho_{i_{a}}\rho_{j_{b}}}\;c_{i_{a}j_{b}}(|\mathbf{r}-\mathbf{r}'|)\,\Big],\label{K-C}
\end{equation}
where $\mathbf{V}$ can be identified (approximately) with the inverse
of the intramolecular correlation matrix $\mathbf{W}$ of RISM, $\mathbf{V}\approx\mathbf{W}^{-1}$.

Using the decomposition in Eq.~(\ref{cBMR}), the short-range (reference
+ short-range bridge) and mesoscopic bridge contributions are separated:
\[
\mathcal{C}_{i_{a}j_{b}}(r)=\mathcal{C}_{i_{a}j_{b}}^{(V)}(r)+\rho_{i_{a}}\rho_{j_{b}}\,b_{i_{a}j_{b}}^{(MR)}(r),
\]
with 
\[
\mathcal{C}_{i_{a}j_{b}}^{(V)}(r)=\sqrt{\rho_{i_{a}}\rho_{j_{b}}}\left[V_{i_{a}j_{b}}(r)-\sqrt{\rho_{i_{a}}\rho_{j_{b}}}\left(c_{i_{a}j_{b}}^{(R)}(r)+b_{i_{a}j_{b}}^{(SR)}(r)\right)\right].
\]
Transforming Eq.~(\ref{F-IET}) to Fourier space, 
\begin{equation}
\begin{aligned}\beta F[\{\rho_{i_{a}}\}]\;=\;\beta F_{0} & \;-\;\frac{1}{2(2\pi)^{3}}\int d\mathbf{k}\int d\mathbf{k}'\;\boldsymbol{\delta\rho}(\mathbf{k})\,\mathbf{C}^{(V)}(|\mathbf{k}+\mathbf{k}'|)\,\boldsymbol{\delta\rho}(-\mathbf{k}')\\
 & \;-\;\frac{1}{2(2\pi)^{3}}\int d\mathbf{k}\int d\mathbf{k}'\;\boldsymbol{\delta\rho}(\mathbf{k})\,\mathbf{B}^{(MR)}(|\mathbf{k}+\mathbf{k}'|)\,\boldsymbol{\delta\rho}(-\mathbf{k}'),
\end{aligned}
\label{F-FT}
\end{equation}
where the superscript “$V$” is stands for ``vacuum'': the first
(short-range) contribution $\mathbf{C}^{(V)}$ plays the role of a
\emph{background} (or “vacuum”) set by molecular details below the
ultraviolet cutoff $\Lambda$, while the second term $\mathbf{B}^{(MR)}$
collects the mesoscopic bridge contributions responsible for domain-scale
structure. In the field-theoretic analogy, $\mathbf{B}^{(MR)}$ is
the microscopic origin of the kernel $\Gamma(k)$ in Eq.~(\ref{FT-expan2}).
Field theories entirely neglect the $\mathbf{C}^{(V)}$ contribution
because the underlying physics is either neglected, such as in soft
matter theories where small solvent and some solute molecules are
ignored, or unknown such as in quantum field theories since the physics
of the vacuum is unknown.

This correspondence provides a microscopic basis for the mesoscopic
TS kernel and motivates the emergence of a finite-$k$ instability
from the site--site description. In the next sections, this conjecture
is supported by analyzing correlation functions of aqueous alkylamine
mixtures from molecular simulations\cite{our_simu_amin} together
with X-ray scattering data\cite{our_expt_amin}.

\section{Illustration with aqueous hexylamine mixtures}

Detailed results from x-ray experiments and computer simulations have
been reported in Refs.\cite{our_expt_amin,our_simu_amin}. Herein,
aqueous hexylamine and aqueous octylamine mixtures are considered,
with appropriate alkyl chain length to make these alkylamine act as
surfactant-like molecules. It is not sufficient, however, that such
molecules have a polar head and an sufficiently long alkyl tail in
order to create micro-segregation. For instance, while octanol has
a similar polar head/alkyl tail structure, aqueous octanol mixture
are immiscible for a large range of solute concentration. The key
to miscibility is that the polar head and water molecule ``anchor''
within each other. In the case of the amine head group, it is the
two hydrogen which allow to ``anchor'' water, or water to ``mingle''
with the amine head groups\cite{2019_Propylamine2}. This is illustrated
through the fact that computer simulations of various models show
segregated water domains to be surrounded by nitrogen atoms, which
in turn stabilize these water nano-domains. 

\begin{figure}[H]
\centering
\includegraphics[scale=0.35]{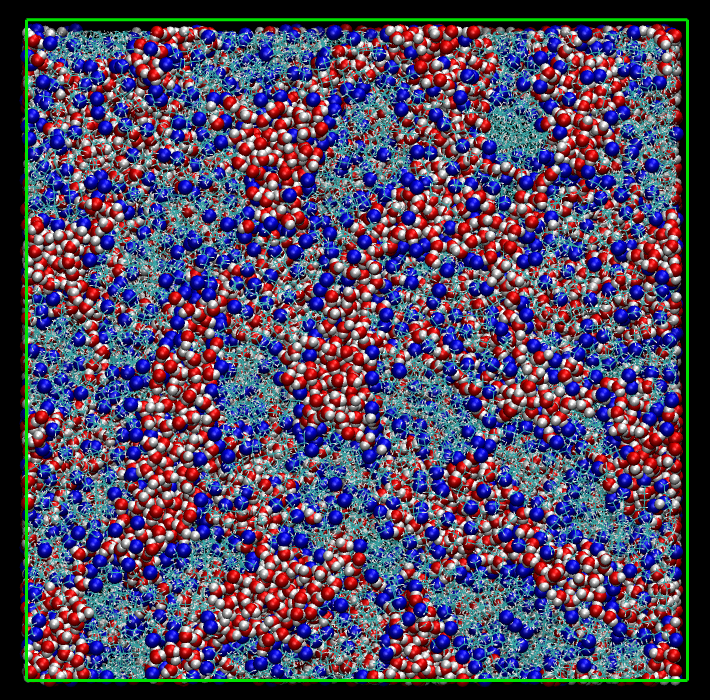}

\caption{Snapshot of the 30\% aqueous-hexylamine mixture illustrating the water
domains surrounded by nitrogen atoms which delimitate them from the
alkyl ``bath''.}

\label{FigSNAP}
\end{figure}

This ``anchoring''/''mingling'' physics does not occur in the
case of aqueous octanol for instance, which is the main reason why
alkanols demix from water starting from butanol, while butylamine
is fully miscible with water, and higher alkylamines mix above 20\%
solute content. For lower concentration, there is no sufficient amine
head groups to stabilize the larger water nano-domains.

\subsection{Teubner-Strey fitting of x-ray spectra}

First, the fact that higher aqueous-alkylamine behave as micro-emulsions
is demonstrated  by fitting the x-ray scattering pre-peaks with the
TS formula of Eq.(\ref{I-TS}).

\begin{figure}[H]
\centering
\includegraphics[scale=0.35]{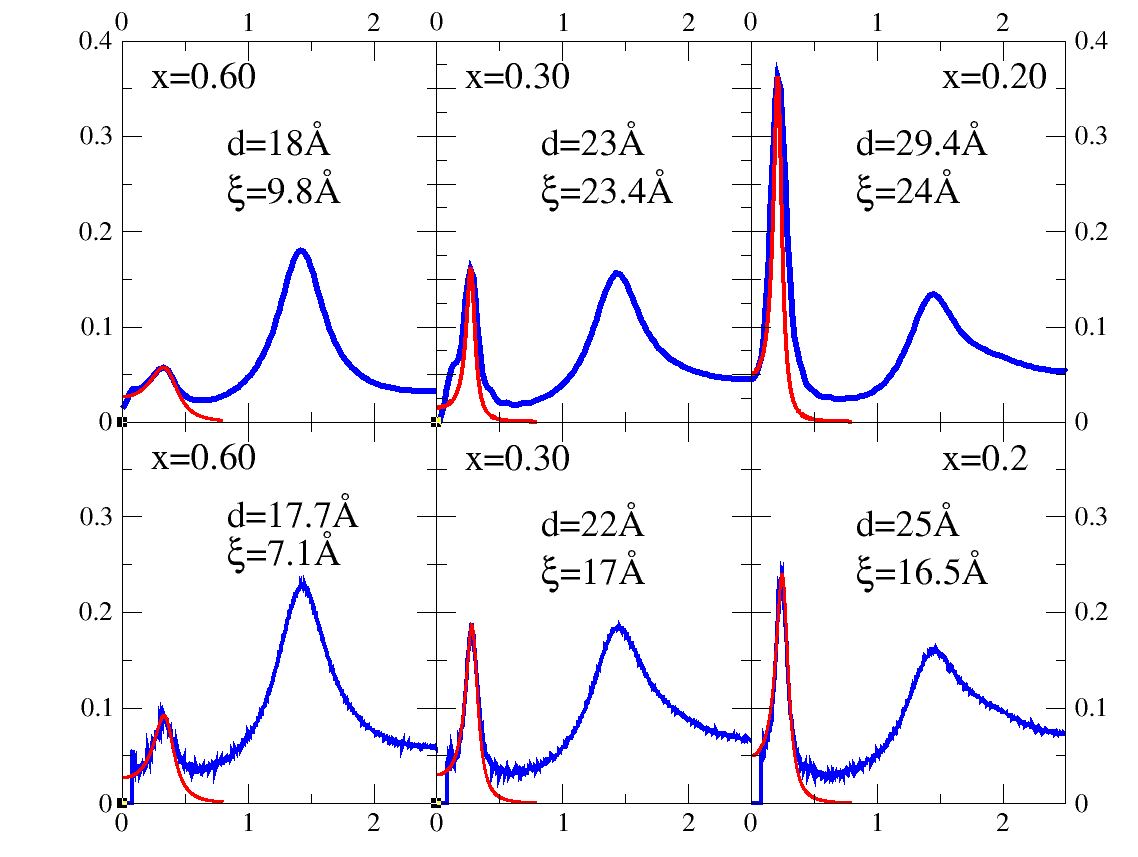}

\caption{Illustration of the TS fitting of the scattering pre-peaks for aqueous-hexylamine,
for various hexylamine mole fractions x shown in each panels. The
x-ray spectra are shown in blue lines, and the TS fitting curves in
red lines. The lower panels are for the experimental data\cite{our_expt_amin}
and the upper panels for the simulation data \cite{our_simu_amin}.
The domain sizes $d$ and OZ correlation lengths $\xi$ are given
in each panel.}

\label{Fig.TS-HEX}
\end{figure}

Fig.\ref{Fig.TS-HEX} shows how the scattering pre-peaks of aqueous
hexylamine mixtures (blue curves) are nicely fitted by the TS formula
(red curves), both for the experimental results (lower panels) and
spectra computed from simulations (upper panels), and for various
hexylamine mole fractions $x$. The values for the domain size $d$
and the OZ correlation length $\xi$ , as extracted from the fits
according to Eqs.(\ref{d-TS},\ref{xi-TS}), are indicated in each
panels. What is remarkable is that the small-$k$ left part of the
SPP is better fitted than the large-$k$ side. This is because the
TS fit is suited for small-$k$, and also because hexylamine is more
of a large molecule than a true surfactant molecule, which is why
the large-$k$ side, closer to distances comparable to atom sizes,
as described by the main peaks around $k_{MP}\approx1.5\mathring{A^{-1}}$,
which is roughly $r\approx2\pi/k_{MP}\approx4\mathring{A}$. Interestingly,
this is more true of the simulation spectra than the experimental
ones, which is probably a problem of the force field model and the
ability to describe entirely the water-solute mingling processes. 

It is also very interesting that the domain size parameter $d$ from
the fit is very close the distance corresponding to the pre-peak positions
$k_{PP}$:
\[
d\approx\frac{2\pi}{k_{PP}}
\]
which confirms that both quantities are related.

It can be seen, however, that the domain sizes and correlation lengths
are in overall good agreement between the experimental and calculated
spectra. This agreement is excellent for large amine content, and
the domain sizes are also in better overall agreement than the correlation
lengths. This is because fluctuations are notoriously difficult to
capture with force field models, as when computing phase boundaries\cite{PhaseDiagSIMU}.

\begin{figure}[H]
\centering
\includegraphics[scale=0.35]{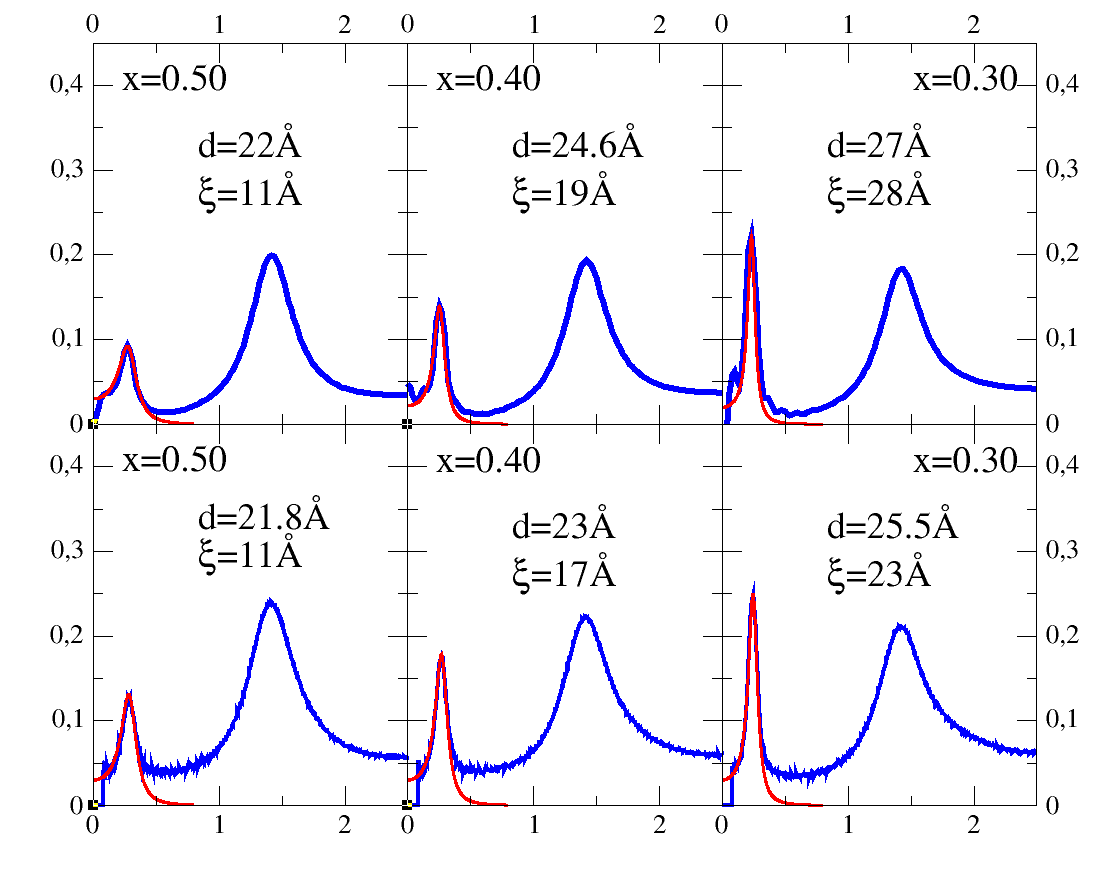}

\caption{Illustration of the TS fitting of the scattering pre-peaks for aqueous-octylamine,
for various hexylamine mole fractions x shown in each panels. The
line and other details are as in Fig.\ref{Fig.TS-HEX}}

\label{Fig.TS-OCT}
\end{figure}

A similar fitting has been conducted for aqueous octylamine, as illustrated
in Fig.\ref{Fig.TS-OCT}. In this case, both the experimental and
calculated spectra are well fitted from the small and large $k$ sides.
It indicates that octylamine might be closer to a surfactant, principally
in view of the alkyl tail length.

The agreements for the $d$ and $\xi$ parameters are comparable to
the case of hexylamine.

It can be seen that the domain size parameters and correlation lengths
increase with increasing water content (decrease with amine mole fraction
$x$) when the water nano-domain become larger. The alkyl size dependence
is less clear. Furthermore, $\xi$ increases faster than $d$ when
the water content is increased, which is consistent with the fact
that $\xi$ should diverge at the spinodal of the phase separation
at small amine mole fractions.

\subsection{Teubner-strey fitting of the tail of the correlation functions}

In Ref.\cite{our_simu_amin}, the large distance tails of the correlation
functions were studied, and it was shown that these showed large scale
oscillations which correspond with the mean segregated domain size.
These were very small amplitude oscillations, whose full decay could
not be observed within the simulation box sizes (around $L\approx100\mathring{A}$,
corresponding to $N=16000$ molecules). An attempt to complement these
oscillatory behaviour by using a functional form that is compatible
with the $k$-space form of TS behaviour in Eq.(\ref{I-TS}) as well
as with the related r-space form Eq.(\ref{I(r)}), is accomplished
with with the following expression.
\begin{equation}
\lim_{r\geq R_{c}}g_{ij}(r)=g_{ij}(R_{c})\exp(-(r-R_{c})/\xi)\sin(\frac{2\pi r}{d})\label{gr-TS}
\end{equation}
where $R_{c}=L/2$ corresponds to half box length, until which the
correlation functions are generally evaluated in computer simulations. 

\begin{figure}[H]
\centering
\includegraphics[scale=0.35]{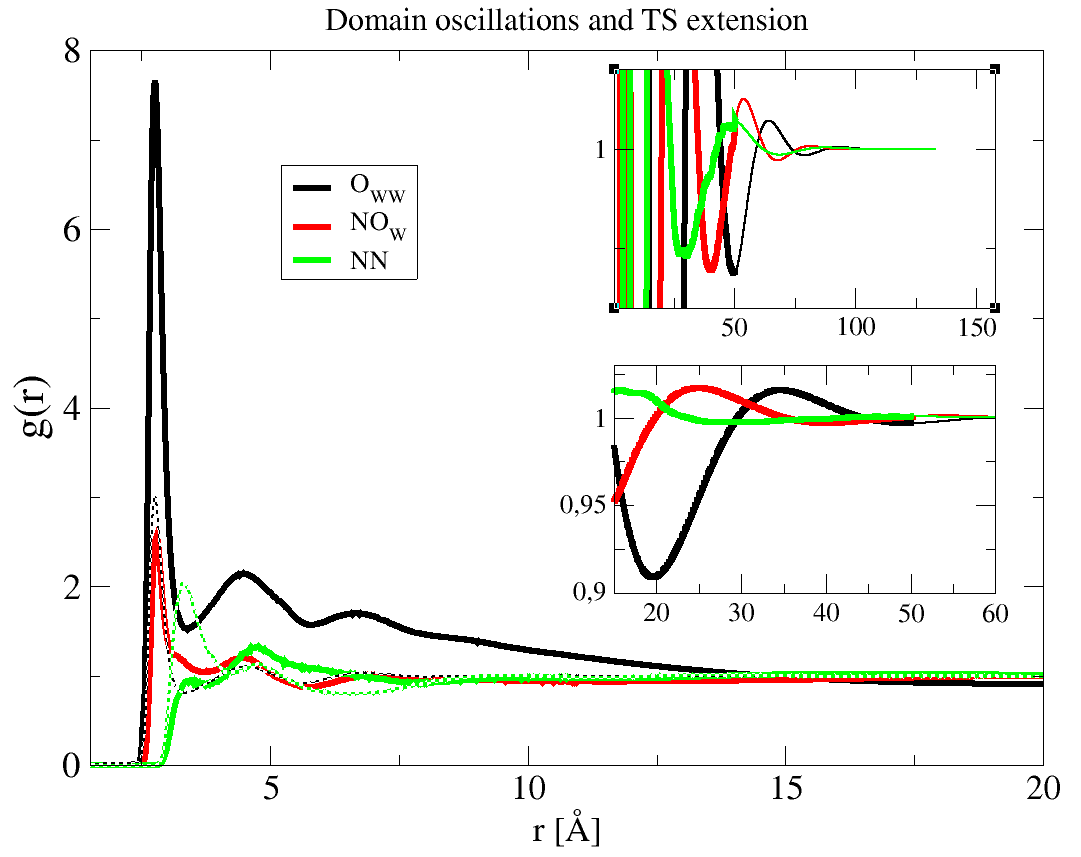}

\caption{Typical atom-atom correlation functions for the 20\% aqueous-hexylamine
mixture, illustrating the short (main panel), medium (lower inset)
and long/meso (upper inset) range behaviour. The neat liquid correlation
functions are shown in dashed lines in the main panel (black for water
and green for hexylamine). The thin lines in the upper inset are the
TS fitting extension of Eq.(\ref{gr-TS}).}

\label{Fig-grTS-hex}
\end{figure}

Fig.\ref{Fig-grTS-hex} illustrates this TS extension fitting of the
$g(r)$ in the case of 20\% aqueous-hexylamine mixtures. Three typical
atom-atom correlation functions are selected, namely $g_{O_{W}O_{W}}(r)$,
$g_{NO_{w}}(r)$ and $g_{NN}(r)$. The short range parts are shown
in the main panel, where the atom-atom correlation with the typical
atom-size oscillations can be seen. Just at the rim of the figure,
for distances about $18-20\mathring{A}$ , the start of the domain
oscillations can be observed, the latter which are highlighted in
the lower inset for distances up to $50\mathring{A}$, and correspond
to half box length for this. Such domain oscillations can be captured
only by allowing sufficiently large system sizes (herein $N=16000$
molecules have been used \cite{our_simu_amin}). If the system size
was limited to N=2000 molecules, the half box size would only cover
distances up to $20\mathring{A}$, which is barely sufficient to observe
the first domain oscillation. Domain oscillation magnitudes are about
one order of magnitude smaller than the atom-atom correlation oscillations.

The TS extension from Eq.(\ref{gr-TS}) can be seen at the rim of
this inset for $r>50\mathring{A}$. This extension is fully illustrated
in the upper inset, where one observes how the fitting smoothly and
naturally extends the domain oscillations above the half-box size
for distances greater than $50\mathring{A}$. The parameters $d$
and $\xi$ are taken from the TS fits of the pre-peak behaviour in
Fig.\ref{Fig.TS-HEX}. In the accompanying SI document, Fig.SI.1 shows
the full range of the TS fitn and how the molecular range and field
theoretic ranges can be separated out.

\begin{figure}[H]
\centering
\includegraphics[scale=0.35]{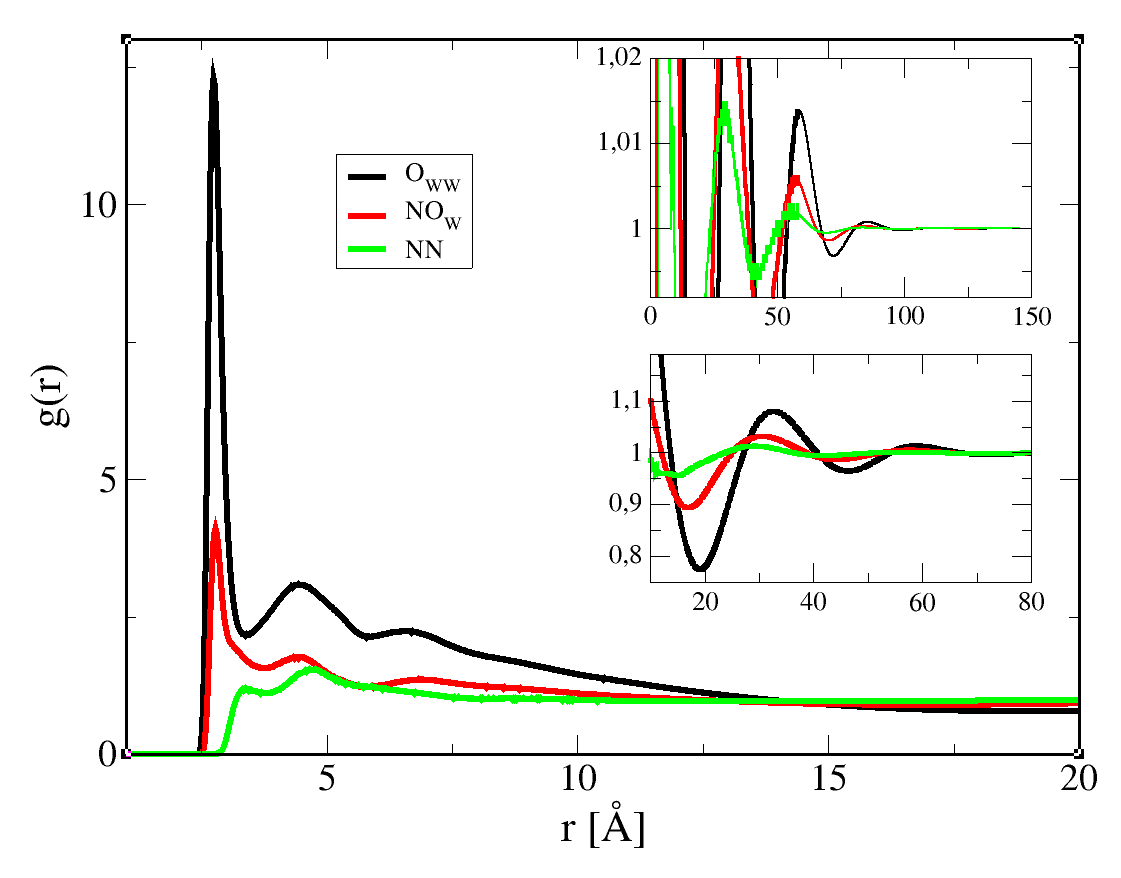}

\caption{Typical atom-atom correlation functions for the 30\% aqueous-octylamine
mixture. The plot details are as in Fig.\ref{Fig-grTS-hex}}

\label{Fig-grTS-OCT}
\end{figure}

A similar fit for the case of 30\% aqueous-octylamine is shown in
Fig.\ref{Fig-grTS-OCT}. In this case, the domain oscillations are
more sustained, and the TS extension fit is better defined.

\begin{figure}[H]
\centering
\includegraphics[scale=0.35]{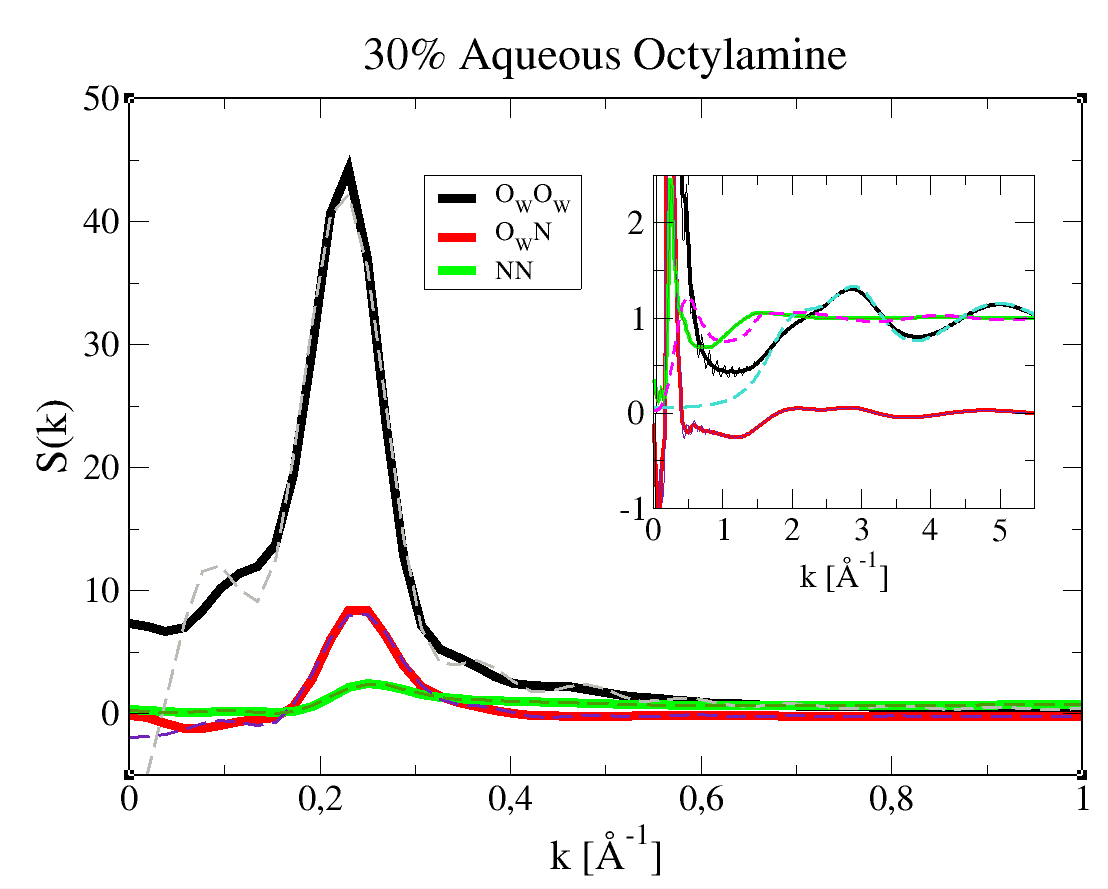}

\caption{Typical atom-atom structure factors $S_{i_{a}j_{B}}(k)$ for the 30\%
aqueous-octylamine mixture. The main panel is a zoom over the pre-peak
region, while the inset shows the main peak region. Thick curves show
the S$(k)$ from the original $g(r)$ obtained from computer simulations.
The thinner curves are the Fourier transforms of the TS-extended $g(r)$
shown in Fig.\ref{Fig-grTS-OCT}.}

\label{FigTS-SK}
\end{figure}

It is equally interesting to examine the structure factors corresponding
to the $g(r)$ shown in Figs.(\ref{Fig-grTS-hex},\ref{Fig-grTS-OCT}).
In Fig.\ref{FigTS-SK} the atom-atom structure factors are shown for
the case of 30\% aqueous-octylamine, corresponding to the $g(r)$
shown in Fig.\ref{Fig-grTS-OCT}. The main panel shows a zoom over
the pre-peaks parts. The original data is shown in thick curves, while
those corresponding to the TS-extension are shown in thinner lines.
It can be seen that, except for the very small k below $k\approx0.2\mathring{A}^{-1}$,
the two curves exactly superpose, confirming that the TS-extension
provides a better $k=0$ behaviour than the S(k) corresponding to
the truncated $g(r)$ of the original data. The original data is seen
to generate small spurious oscillations, more visible in the inset.

The inset shows the main peak region, together with the neat liquids
$S(k)$, shown in dashed cyan for water and magenta for octylamine.
It is seen that the structure of the water domain is nearly identical
to that of bulk water. 

\begin{figure}[H]
\centering
\includegraphics[scale=0.35]{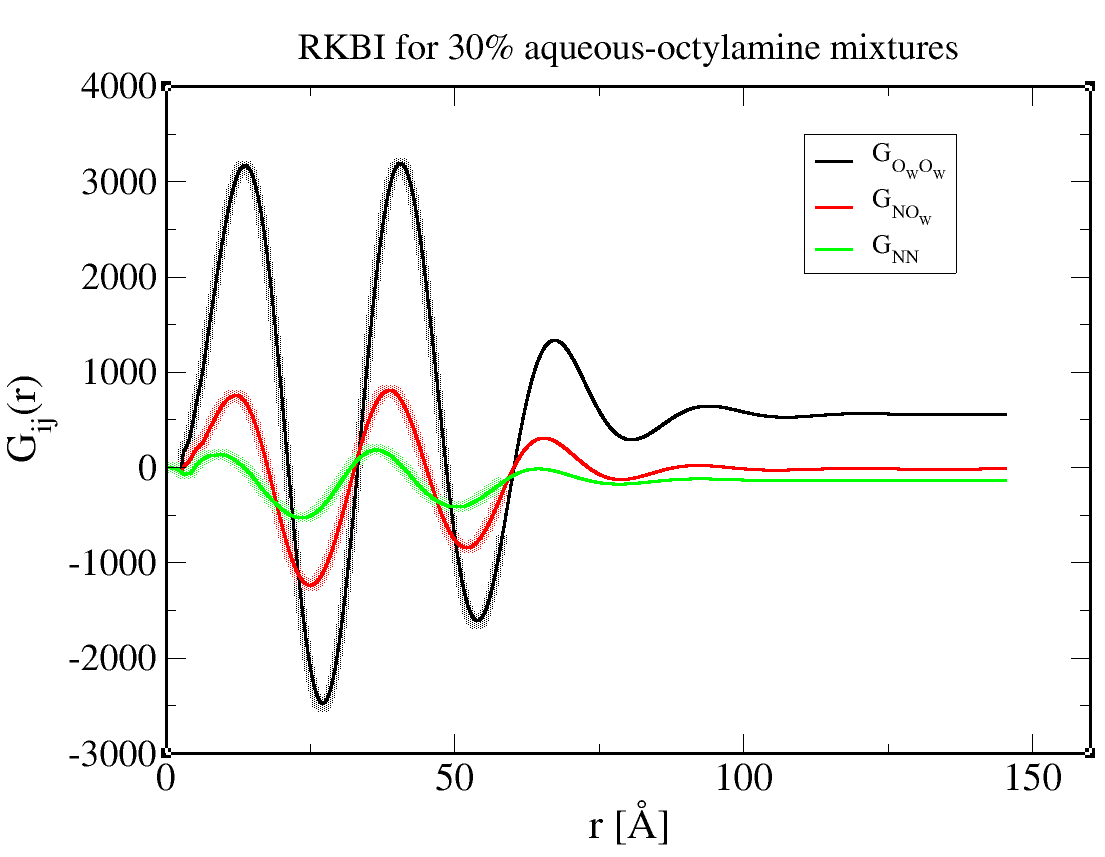}

\caption{Running Kirkwoof-buff integrals for the 30\% aqueous-octylamine mixtures.
The thick lines represent RKBI of the $g(r)$ from simulation data,
and the thinner lines the RKBI from the TS-extended $g(r)$.}

\label{FigRKBI}
\end{figure}

In order to better enforce the quality of the TS-extension, the running
Kirkwood-Buff integrals (RKBI) defined by \cite{CamelBackKBI}, ae
shown in Fig.\ref{FigRKBI}
\begin{equation}
G_{i_{a}j_{b}}(r)=4\pi\int_{0}^{r}dss^{2}\left[g_{i_{a}j_{b}}(r)-1\right]\label{RKBI}
\end{equation}
whose $r\rightarrow\infty$ limit are the Kirkwood-Buff integrals
(KBI). Fig.\ref{FigRKBI} demonstrates the smooth continuation of
the domain oscillations towards well defined horizontal asymptotes
which define the KBI.

It is obvious that such asymptotes cannot be obtained from the current
simulations, since the corresponding data stops at $r\approx60\mathring{A}$,
as indicated by the thick lines. In this region, the domain oscillations
are still important and cannot help decide the KBI values. The TS-extension
allows unambiguous determination of the KBI. It also confirms that
separation between the inner range and meso-range correlations, which
is conjectured in Section \ref{sec:Liquid-state-statistical}

\begin{figure}[H]
\centering
\includegraphics[scale=0.35]{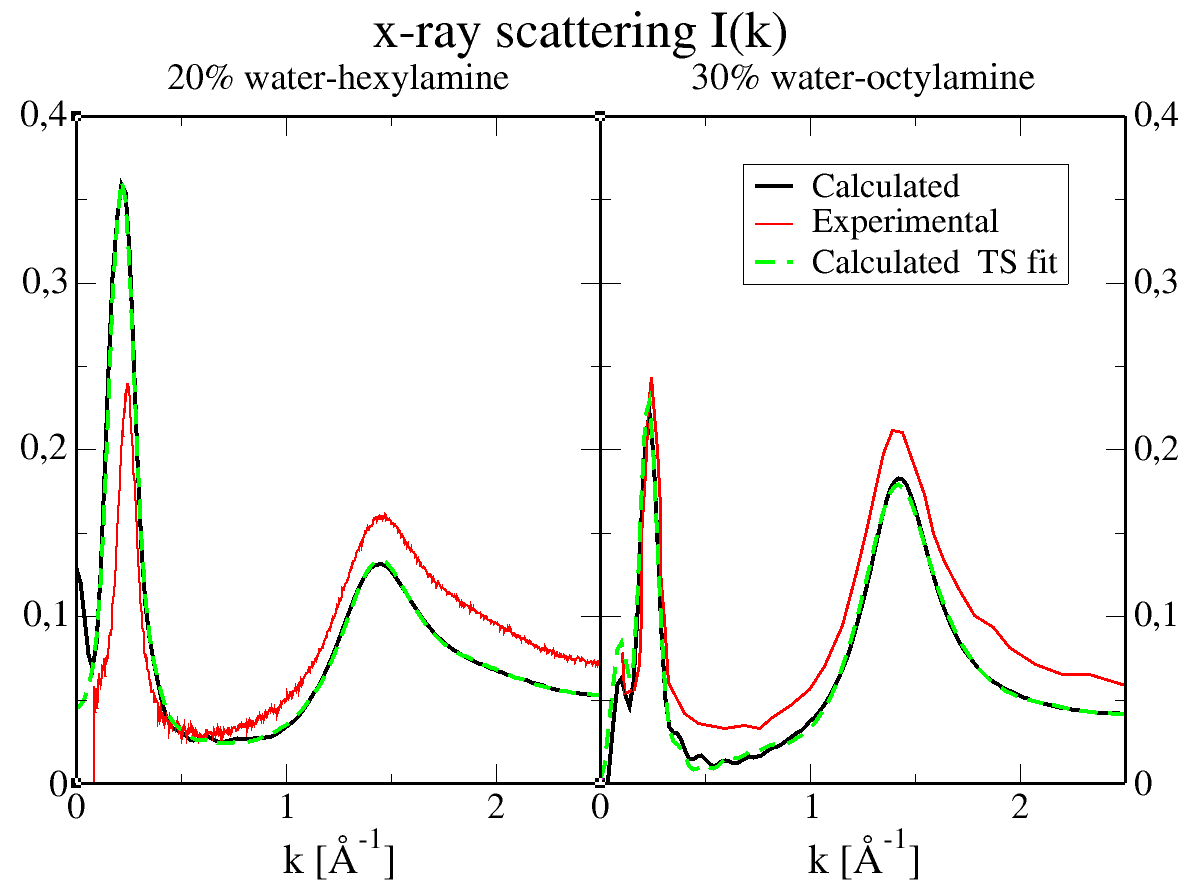}

\caption{x-ray scattering intensities comparing the experimental data (red
curves) to the calculated ones, one from the g(r) obtained from the
simulations (black lines) and that from the TS-extended g(r) (dashed
green lines). Left panel is for 20\% aqueous hexylamine and right
panel for 30\% aqueous octylamine.}

\label{FigSAXS}
\end{figure}

Finally, in order to confirm that the TS extension does not alter
the prediction of the scattering intensities, Fig.\ref{FigSAXS} shows
the x-ray spectra obtained from the original g(r) from the simulations
and that obtained from the TS-extension of each of the atom-atom g(r).
These are shown in dashed curves. It can be seen that the superposition
of both curves in nearly exact, except from the small numerical artifacts
from the Fourier transforms of the original g(r) which do not converge
to 1 at the end of the half-box. 

The figure equally shows the type of agreement with the experimental
data one can obtain from model computer simulations, which was previously
studied in Refs.\cite{our_simu_amin,our_expt_amin}. The agreement
tends to be better for the longer octylamine. This enforces the conjecture
proposed in Section \ref{sec:Liquid-state-statistical}, since it
shows that systems closer to the micro-emulsion are better described
by model simulations, precisely because the meso-range structure decouples
better from the small range atom/molecular range, which has been called
herein called the ``Vacuum'' range.

The SI document shows how the TS expression, while describing perfectly
the meso-physics under the pre-peak, cannot account for the molecular
details. Fig.SI.2 shows how $I(k)$ is the result of huge cancellations
between the partial scattering intensities from water, octylamine
and cross species contributions. Similarly, Fig.SI.3 illustrates the
partial atom-astom structure factors contain all the molecular details,
including those which contribute to the pre-peaks. These details can
only be captured by a fully microscopic theory.

\section{Discussion\protect}\label{sec:Discussion}

The central hypothesis of this work---that the mesoscopic correlations
of soft-matter systems originate from the meso-range component of
the bridge function---is far from trivial. It challenges the long-standing
view that the bridge function is always shorter-ranged than the direct
correlation function \cite{FisherCrit}. This belief is supported
by studies of simple liquids, where $b(r)$ has been extracted from
simulations of Lennard--Jones\cite{bridgeLJ,bridgeLJchapman} or
hard-sphere\cite{LabikBridge} models. In those cases, as well as
in self-consistent closures\cite{bridgeLJVerlet}, the bridge function
indeed appears weak and rapidly decaying.

However, complex liquids displaying self-assembly or domain formation
behave differently. Our conjecture is supported by the limited success
of the hypernetted-chain (HNC) closure, which---though the most complete
integral-equation theory---neglects the bridge function entirely\cite{hncOrig}.
We argue that it is precisely this neglect that prevents HNC from
describing molecular emulsions and soft-matter systems properly: the
missing higher-order terms embedded in the meso-range bridge function
(Eq.~\ref{b12}) are essential to represent collective mesoscopic
organization.

The apparent success of alternative closures such as Percus--Yevick
(PY)\cite{Textbook_Hansen_McDonald} , hybrid mean spherical approximation
(HMSA)\cite{HMSA} , or Kovalenko--Hirata (KH)\cite{hirata2025molecular}
closures has, in retrospect, diverted attention from the fundamental
limitations of HNC. The bridge function remains difficult to extract
reliably from simulations,\cite{BridgeSimuProblem,bridgeSimuMixt}
yet distinguishing between its short-range and mesoscopic parts clarifies
its physical role. The short-range component governs local coordination,
whereas the meso-range component controls domain-scale aggregation
and self-assembly.

Even if short-range structural details remain imperfectly captured,
incorporating an appropriate meso-range bridge term may allow HNC-type
closures to reproduce soft-matter structure quantitatively. The Teubner--Strey
form used here thus appears as a promising kernel for developing explicit
mesoscopic bridge functionals that can be integrated within the framework
of molecular integral-equation theory.

\section{Conclusion\protect}\label{sec:Conclusion}

A theoretical framework has been proposed, which unifies molecular
liquids and soft-matter systems such as micro-emulsions within a single
statistical-mechanical description. The key conjecture is that the
long-range structure of complex liquids originates from mesoscopic
contributions of the bridge function---a hypothesis that departs
from the established assumption of its purely short-ranged character
in simple liquids.

Through the example of aqueous alkylamine mixtures, which exhibit
persistent micro-heterogeneity, it was shown that domain-scale oscillations
in the correlation functions stem from a Teubner--Strey--type kernel.
This mesoscopic component decouples from the local, atomistic correlations
and provides the microscopic foundation of the Landau--Ginzburg--de
Gennes field theory used to describe modulated phases.

The resulting connection between molecular-scale statistical mechanics
and mesoscopic field theories opens the way for integral-equation
approaches to explore more complex soft-matter systems. Such a unified
description offers a valuable theoretical complement to computer simulations,
bridging the gap between microscopic interactions and emergent supramolecular
organization.

\section*{Data availability}

The data used in this work is available upon request to the author.

\section*{Supporting Information}

The Supporting Information document contains information on the short
coming of field theoretical approaches when it concern molecular details.

\section*{Acknowledgments}

The author thanks Bernarda Lovrinčević, Martina Požar and Christian
Sternemann for insightful comments.

This work was supported by the French-German collaborations PROCOPE
(50951YA), \emph{Analysis of the molecular coherence in the self-assembly
process: experiment and theory}.

\bibliographystyle{jpcb_final.bst}
\bibliography{wamines}

\end{document}